# Heteroepitaxial Growth of High-Mobility Ge-Doped β-$Ga_2O_3$ Films on Sapphire Substrate by Low-Pressure CVD

Ahmed Ibreljic, Saleh Ahmed Khan, Sourav Sarker, Ibrahim Isah, Stephen Margiotta, Michael T. Davenport, Stephen Lam, and Anhar Bhuiyan*

**ABSTRACT:** In this work, high-quality Ge-doped $(\bar{2}01)$ β-$Ga_2O_3$ thin films were heteroepitaxially grown on *c*-plane sapphire substrates with offcut angles of 0°, 2°, 6°, and 8° using low-pressure chemical vapor deposition (LPCVD). Increasing the sapphire offcut promoted step-flow growth, resulting in improved terrace alignment, reduced surface roughness, and enhanced crystalline quality. Phase-pure monoclinic β-$Ga_2O_3$ with strong $(\bar{2}01)$ preferential orientation was confirmed by X-ray diffraction and Raman spectroscopy, while X-ray photoelectron spectroscopy revealed a near-stoichiometric composition with an O/Ga ratio of 1.48. Electrical transport properties exhibited a strong dependence on the substrate offcut angle, with room-temperature Hall mobility increasing from 15 to 117 $cm^2/V\cdot s$ as the offcut angle increased from 0° to 6°, across carrier concentrations spanning $1.43 \times 10^{17}$ to $2.75 \times 10^{18}$ $cm^{-3}$. The 6° offcut sample achieved a room-temperature mobility of 117 $cm^2/V\cdot s$ at a carrier concentration of $1.43 \times 10^{17}$ $cm^{-3}$ and a peak low-temperature mobility of 337 $cm^2/V\cdot s$ at 128 K with a carrier concentration of $8.96 \times 10^{16}$ $cm^{-3}$, representing the highest reported room-temperature and low-temperature mobilities for Ge-doped β-$Ga_2O_3$ films grown on sapphire substrates. Carrier concentration and mobility data were analyzed using charge-neutrality and Boltzmann transport models incorporating donor activation together with polar optical phonon, ionized impurity, neutral impurity, acoustic deformation potential, and dislocation scattering mechanisms. The fitting revealed shallow donor activation energies of 12.5–19 meV, a deeper donor level at 80 meV, low acceptor compensation ($<5 \times 10^{15}$ $cm^{-3}$), and threading dislocation densities on the order of $10^9$ $cm^{-2}$. These results establish sapphire offcut engineering as an effective pathway for achieving high-mobility, LPCVD-grown Ge-doped β-$Ga_2O_3$ heteroepitaxy on scalable foreign substrates, which is critical for the development of high-performance β-$Ga_2O_3$ power devices.

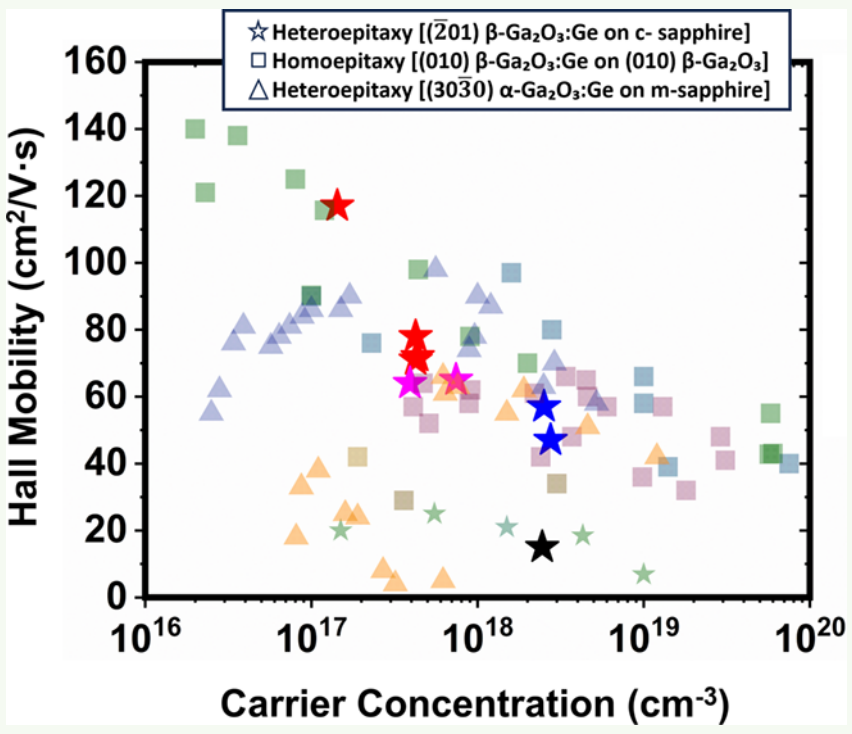

**KEYWORDS:** *ultra-wide bandgap semiconductor, low pressure chemical vapor deposition LPCVD, β-$Ga_2O_3$, sapphire substrate, Ge doping*

## 1. INTRODUCTION

β-$Ga_2O_3$ has emerged as a promising ultrawide bandgap semiconductor for next-generation high-power electronics due to its large bandgap (4.8–4.9 eV), high critical breakdown field (6–8 MV/cm), and the availability of large-area melt-grown native substrates with excellent crystalline quality.[1] Rapid progress has been achieved in recent years in both β-$Ga_2O_3$ epitaxial growth and device demonstration. High-quality β-$Ga_2O_3$ films have been successfully developed using several advanced growth techniques, including metal–organic chemical vapor deposition (MOCVD),[2−15] halide vapor phase epitaxy (HVPE),[16−19] molecular beam epitaxy (MBE),[20−24] low-pressure chemical vapor deposition (LPCVD),[25−32] pulsed laser deposition (PLD),[33,34] and mist chemical vapor deposition (mist-CVD).[35−37] These approaches have enabled significant advances in crystal quality, doping control, growth rate, surface morphology, and large-area epitaxy on both native and foreign substrates. Concurrently, substantial advances have been made in β-$Ga_2O_3$ device technologies,[38−43] including the demonstration of multi-kilovolt-class Schottky barrier diodes,[44−49] p–n heterojunction diodes,[50−53] and both lateral and vertical field-effect transistors[54−59] with high breakdown voltage and low specific on-resistance. These developments highlight the strong potential of β-$Ga_2O_3$ for next-generation high-power and high-voltage electronics.

Achieving high-performance β-$Ga_2O_3$ power devices critically depends on precise control of n-type doping across a wide concentration range, from lightly doped drift layers below $10^{15}$ $cm^{-3}$ to heavily doped contact layers exceeding $10^{19}$ $cm^{-3}$. Several donor impurities can be incorporated into β-$Ga_2O_3$,

**Received:** May 19, 2026
**Revised:** June 22, 2026
**Accepted:** July 25, 2026
**Published:** August 9, 2026

with Si, Sn, and Ge being the most widely studied. First-principles calculations and experimental transport studies have shown that Si and Ge preferentially occupy tetrahedrally coordinated Ga(I) lattice sites, giving rise to shallow donor levels, whereas Sn preferentially occupies octahedral Ga(II) sites associated with deeper donor levels and reduced room-temperature activation efficiency.[60,61] Among these dopants, Ge offers several potential advantages, including a stronger energetic preference for shallow donor configurations compared to Si, which may enable higher free electron concentrations at comparable doping levels. In addition, the larger ionic radius of $Ge^{4+}$ provides improved lattice matching to Ga sites,[62] reducing local lattice distortion and potentially enhancing crystalline uniformity.

Ge-doped β-$Ga_2O_3$ has been demonstrated by MBE[23] and MOCVD[60] through homoepitaxial growth on native β-$Ga_2O_3$ substrates, achieving room-temperature electron mobilities ranging from ~90 to 140 $cm^2$/V·s over carrier concentrations spanning the low-$10^{16}$ to low-$10^{18}$ $cm^{-3}$ range. In plasma-assisted MBE-grown films,[23] electron mobilities up to 97 $cm^2$/V·s were reported at carrier concentrations of $1.6 \times 10^{18}$ $cm^{-3}$, while heavily doped films with carrier concentrations approaching $10^{20}$ $cm^{-3}$ exhibited mobilities in the range of 38–40 $cm^2$/V·s. Similarly, MOCVD-grown Ge-doped β-$Ga_2O_3$ films demonstrated free carrier concentrations ranging from $\sim 2 \times 10^{16}$ to $3 \times 10^{20}$ $cm^{-3}$, with corresponding mobilities varying from ~140 to 38 $cm^2$/V·s.[60] Although these studies demonstrate the promise of Ge as an effective n-type dopant for β-$Ga_2O_3$, most high-mobility results have been achieved through homoepitaxial growth on expensive native substrates, while scalable heteroepitaxial approaches on low-cost foreign substrates remain largely underexplored.

Among the available foreign substrates, *c*-plane sapphire is particularly attractive for large-area β-$Ga_2O_3$ heteroepitaxy because of its low cost, commercial wafer scalability, thermal stability, and compatibility with existing semiconductor manufacturing infrastructure. However, heteroepitaxial growth of β-$Ga_2O_3$ on on-axis sapphire commonly results in rotational domains, increased defect densities, and degraded electrical transport properties. Intentional sapphire offcut has previously been shown to suppress rotational variants and promote step-flow growth in Si-doped β-$Ga_2O_3$ films,[27] leading to substantial improvements in crystalline quality and carrier mobility. Nevertheless, the influence of substrate offcut on Ge incorporation, structural evolution, and carrier transport in Ge-doped β-$Ga_2O_3$ heteroepitaxy remains insufficiently understood.

In this context, LPCVD presents an attractive alternative growth approach for scalable β-$Ga_2O_3$ heteroepitaxy. Unlike MOCVD and HVPE, which rely on metal–organic precursors and chloride-based chemistries, respectively, LPCVD utilizes elemental metallic gallium and oxygen as growth precursors. This simplified chemistry potentially reduces precursor-related impurity incorporation and enables a cleaner growth environment. LPCVD has already demonstrated successful homoepitaxial growth on native β-$Ga_2O_3$ substrates as well as heteroepitaxial growth on sapphire. Previous studies on LPCVD-grown Ge-doped β-$Ga_2O_3$ films reported room-temperature mobilities typically in the range of ~20 to 43 $cm^2$/V·s over carrier concentrations of $1.5 \times 10^{17}$ to $3 \times 10^{18}$ $cm^{-3}$,[63] indicating that optimization of heteroepitaxial growth conditions is still needed. At the same time, our prior work on LPCVD-grown Si-doped β-$Ga_2O_3$ heteroepitaxy on sapphire demonstrated room-temperature mobilities as high as 149 $cm^2$/V·s, together with strong rectifying device characteristics,[29,31] highlighting the significant potential of LPCVD for high-quality β-$Ga_2O_3$ heteroepitaxy.

Motivated by these considerations, in this work, we systematically investigate LPCVD-grown Ge-doped β-$Ga_2O_3$ films on *c*-plane sapphire substrates with offcut angles of 0°, 2°, 6°, and 8°. A comprehensive study was carried out to evaluate the influence of sapphire offcut on epitaxial growth behavior, Ge incorporation, crystalline quality, surface morphology, electrical transport, and temperature-dependent carrier dynamics. Structural and morphological evolution associated with substrate miscut was examined to understand the suppression of rotational domains and the transition toward improved step-flow growth. Hall-effect measurements were further employed to establish correlations between substrate offcut, carrier concentration, and electron mobility, while temperature-dependent transport analysis was used to probe donor activation and carrier scattering behavior. Through this systematic offcut-dependent investigation, the present work establishes the critical role of sapphire offcut engineering in optimizing LPCVD-grown Ge-doped β-$Ga_2O_3$ heteroepitaxy and provides important insights for the scalable growth of high-quality β-$Ga_2O_3$ films on low-cost foreign substrates.

## 2. EXPERIMENTAL DETAILS

Ge-doped ($\bar{2}01$) β-$Ga_2O_3$ thin films were heteroepitaxially grown on nominally on-axis (0°), 2°, 6°, and 8° off-axis *c*-plane sapphire substrates using a custom-built LPCVD system. The off-axis sapphire substrates were intentionally miscut toward the $\langle 11\bar{2}0 \rangle$ crystallographic direction. The LPCVD reactor consists of a three-zone furnace (Zones I–III), with all zones maintained at 1100 °C. The Ge and Ga source materials were placed in separate crucibles inside an inner quartz tube. The Ge source was positioned in Zone I, while the Ga source and sapphire substrate were located in Zone II. The Ge crucible was positioned upstream of the Ga source, enabling Ge-containing vapor species generated from the Ge source to be transported downstream into the growth zone by the carrier gas. The Ga-substrate distance was varied from 2.5 to 5.5 cm, while the Ge–Ga source spacing was maintained at 33 cm for all growths, as summarized in Table 1. Prior to growth, the sapphire substrates were sequentially cleaned in acetone, isopropyl alcohol (IPA), and deionized (DI) water, followed by nitrogen drying. No additional surface pretreatment, chemical treatment, or high-temperature annealing was performed prior to growth. High-purity gallium pellets (99.99999%), metallic Ge pellets (99.9999%), oxygen gas (99.999%), and argon gas (99.9999%) were used as the Ga source, the n-type dopant source, the oxidant, and the carrier gas, respectively. The Ge-doped β-$Ga_2O_3$ films were grown at a substrate temperature of 1100 °C under a chamber pressure of ~1.5 Torr with Ar and $O_2$ flow rates of 100 and 20 sccm, respectively. The Ge incorporation was controlled by varying the Ge source loading while maintaining identical growth temperatures, pressures, and gas flow rates. The relative Ge loading was quantified using the Ge/(Ge + Ga) weight percentage, which was varied from 0.05 to 0.60 wt % across the investigated samples. For samples grown under identical source geometry conditions, increasing the Ge loading increases the Ge vapor flux reaching the substrate, thereby enabling systematic control of donor incorporation. The corresponding Ge/(Ge + Ga) weight percentages, source geometry parameters, and electrical properties are summarized in Table 1.

The surface morphology of the as-grown films was characterized using field-emission scanning electron microscopy (FESEM) (JEOL JSM-7401F) and atomic force microscopy (AFM) (Park XE-100). Film thicknesses were estimated from cross-sectional FESEM images. The crystalline structure, orientation, and quality were evaluated using high-resolution X-ray diffraction (XRD) with a Rigaku SmartLab system equipped with a Cu Kα radiation source (λ = 1.5418 Å). Raman spectroscopy was carried out using a Horiba LabRAM Evolution Raman spectrometer with a 532 nm excitation wavelength to assess phase purity and lattice vibrational modes. X-ray photoelectron spectroscopy (XPS) measurements were carried out using a Thermo Scientific XPS spectrometer with a monochromatized Al Kα X-ray source ($E_{photon}$ = 1486.6 eV) to determine the Ga and O content in the films. Hall-effect measurements were performed using the van der Pauw configuration with an Ecopia HMS-5300 Hall measurement system. Ohmic contacts were formed at the four corners of the samples using indium dots. No additional metallization or post-contact annealing was employed. Measurements were conducted over the temperature range of 80–350 K.

Table 1. Growth and Electrical Transport Properties of Ge-Doped β-$Ga_2O_3$ Films Grown on Sapphire Substrates with Different Offcut Angles

| sample no. | substrate miscut (deg) | Ga-substrate distance (cm) | $\frac{Ge}{Ge+Ga}$ (wt %) | film thickness (μm) | growth rate (μm/h) | carrier concentration ($cm^{-3}$) | Hall mobility($cm^2$/V·s) |
|---|---|---|---|---|---|---|---|
| 1 | 0 | 2.5 | 0.50 | 1.14 | 1.14 | $2.45 \times 10^{18}$ | 15 |
| 2 | 2 | 3.5 | 0.50 | 1.69 | 1.69 | $2.51 \times 10^{18}$ | 57 |
| 3 | 2 | 3.5 | 0.60 | 1.70 | 1.70 | $2.75 \times 10^{18}$ | 47 |
| 4 | 6 | 4.5 | 0.05 | 3.01 | 3.01 | $1.43 \times 10^{17}$ | 117 |
| 5 | 6 | 4.5 | 0.10 | 3.66 | 3.66 | $4.24 \times 10^{17}$ | 78 |
| 6 | 6 | 4.5 | 0.10 | 2.66 | 2.66 | $4.36 \times 10^{17}$ | 72 |
| 7 | 6 | 4.5 | 0.15 | 5.34 | 3.56 | $4.55 \times 10^{17}$ | 71 |
| 8 | 8 | 5.5 | 0.10 | 4.12 | 4.12 | $3.92 \times 10^{17}$ | 64 |
| 9 | 8 | 5.5 | 0.20 | 0.85 | 2.55 | $7.42 \times 10^{17}$ | 65 |

## 3. RESULTS AND DISCUSSION

To investigate the influence of *c*-plane sapphire offcut on the surface morphology of Ge-doped β-$Ga_2O_3$ films, plan-view scanning electron microscopy (SEM) was performed, as shown in Figure 1a–d. The film grown on nominally on-axis (0° offcut) *c*-plane sapphire, shown in Figure 1a, exhibits a high density of irregular pseudo-hexagonal surface features with random lateral orientations and discontinuous morphology. The absence of preferential step guidance on the nominally flat substrate surface promotes multidirectional lateral growth, resulting in fragmented surface structures and limited long-range ordering. With the introduction of a 2° offcut toward the ⟨11$\bar{2}$0⟩ direction, the surface morphology evolves substantially, as shown in Figure 1b. Elongated surface features become preferentially aligned along the offcut direction, indicating the onset of step-flow-assisted growth. The substrate steps introduced by the intentional offcut provide energetically favorable incorporation sites for migrating adatoms, thereby suppressing randomly oriented rotational domains and improving lateral ordering. Further enhancement in surface ordering is observed for the film grown on the 6° offcut sapphire substrate, shown in Figure 1c, where the surface becomes significantly more uniform and continuous with well-aligned terraces extending over larger lateral dimensions. The reduction in domain boundaries and increased terrace continuity suggest stabilization of the step-flow growth regime. For the film grown on the 8° offcut sapphire substrate (Figure 1d), strong long-range terrace alignment is maintained. Compared to the 6° sample, slightly increased variations in terrace width and localized step bunching features are observed, which are commonly associated with the higher step density introduced at larger substrate miscut angles. Nevertheless, the surface morphology remains highly ordered and continuous, indicating that step-flow growth is effectively sustained at both 6° and 8° offcut conditions.

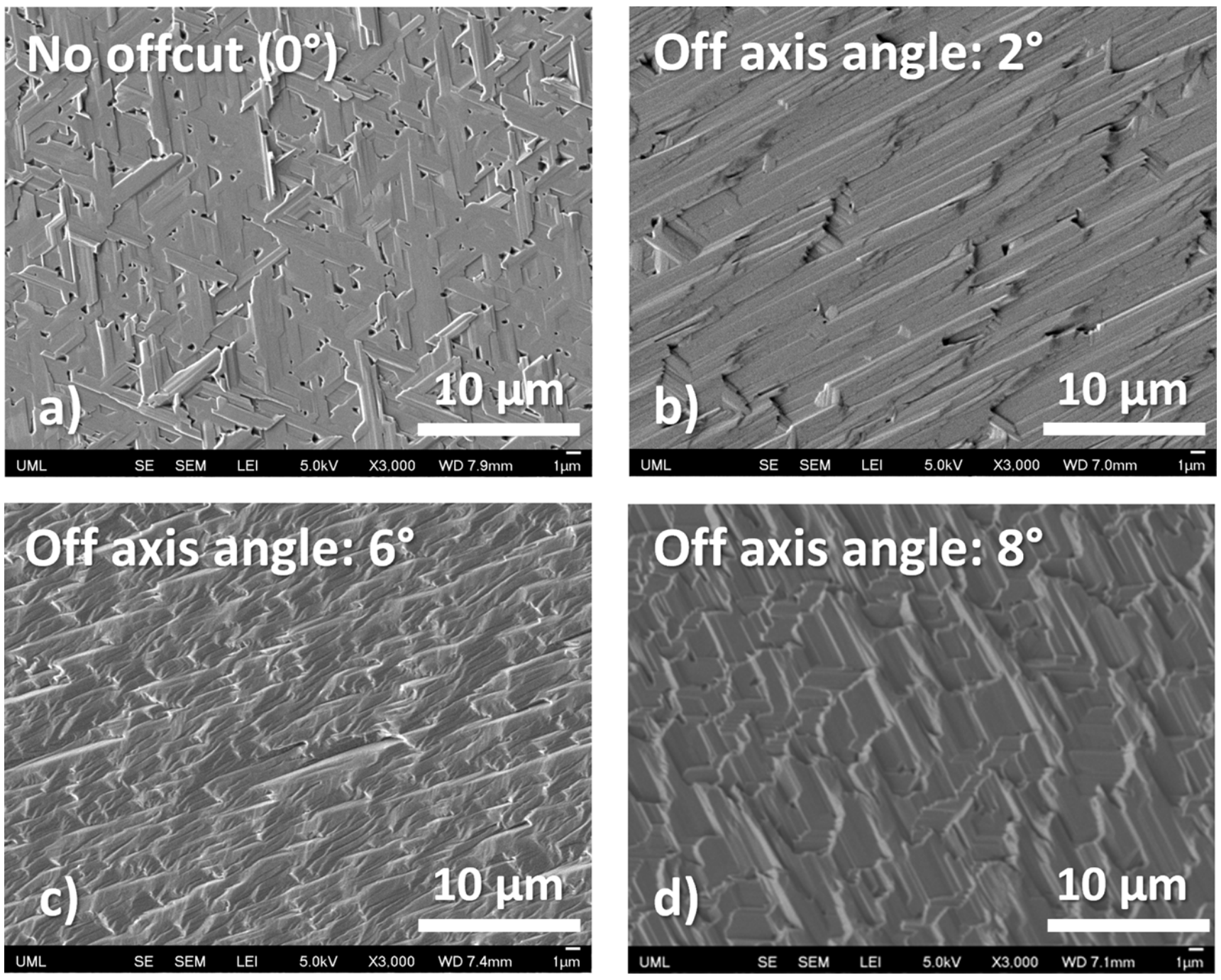

Figure 1. Surface view FESEM images of LPCVD-grown Ge-doped β-$Ga_2O_3$ thin films heteroepitaxially grown on *c*-plane sapphire substrates with different offcut angles: (a) nominally on-axis (0°), (b) 2°, (c) 6°, and (d) 8° offcut.

AFM was further performed to quantitatively evaluate the surface morphology evolution, as shown in Figure 2a–d. The AFM image of the film grown on on-axis sapphire, shown in Figure 2a, reveals irregular terraces and multidirectional step structures consistent with the non-uniform surface morphology observed in the SEM analysis. The corresponding root-mean-square (RMS) surface roughness is measured to be 7.28 nm. With the introduction of a 2° offcut, shown in Figure 2b, the surface morphology transitions toward elongated and more continuous terraces aligned along the offcut direction, further confirming the emergence of step-flow growth. The RMS roughness decreases slightly to 7.05 nm, indicating a modest improvement in surface uniformity while localized step-height fluctuations remain present. For the film grown on the 6° offcut substrate, shown in Figure 2c, the AFM image exhibits well-defined and uniformly aligned terraces with significantly improved lateral continuity and reduced terrace-height fluctuations. This sample exhibits the lowest RMS roughness of 6.84 nm, indicating optimized step-flow growth and enhanced surface coherence. At the higher offcut angle of 8°, shown in Figure 2d, the surface continues to exhibit highly aligned terrace structures with strong directional ordering. The RMS roughness remains low at 6.85 nm, indicating that the higher substrate offcut maintains stable step-flow growth and improved surface smoothness. To assess the statistical significance of the roughness measurements, RMS roughness values were extracted from fifteen independent 2 μm × 2 μm AFM scans obtained from different locations within a larger 30 μm × 30 μm scan area for each sample. The resulting average RMS roughness values were 8.01 ± 0.59, 7.87 ± 0.97, 6.32 ± 1.47, and 6.95 ± 0.48 nm for the 0°, 2°, 6°, and 8° offcut samples, respectively. These values are consistent with the overall trend shown by the representative AFM images in Figure 2. These results demonstrate that increasing the sapphire offcut angle promotes directional terrace alignment and enhances the surface ordering of LPCVD-grown Ge-doped β-$Ga_2O_3$ thin films.

The crystalline orientation and structural quality of the LPCVD-grown Ge-doped β-$Ga_2O_3$ thin films were further investigated using high-resolution XRD, as shown in Figure 3a,b. Figure 3a presents the ω–2θ scan obtained from the film grown on the 6° offcut sapphire substrate. In addition to the sapphire substrate reflections, the diffraction pattern is dominated by the $(\bar{2}01)$, $(\bar{4}02)$, and $(\bar{6}03)$ reflections of monoclinic β-$Ga_2O_3$, confirming phase-pure β-$Ga_2O_3$ growth with strong out-of-plane preferential orientation along the $(\bar{2}01)$ family of planes. No diffraction peaks associated with secondary gallium oxide phases are detected, indicating high phase purity under the present LPCVD growth conditions. To further quantify the crystalline quality, rocking curve measurements were performed on the symmetric $(\bar{4}02)$ reflection for films grown on sapphire substrates with different offcut angles, as shown in Figure 3b. The rocking curve full width at half maximum (FWHM) decreases systematically with increasing offcut angle from 0° to 8°, indicating progressively improved crystalline coherence. This trend correlates strongly with the enhanced terrace alignment and improved surface uniformity observed in the SEM and AFM analyses shown in Figures 1 and 2. Among the investigated samples, the films grown on the 6° and 8° offcut sapphire substrates exhibit the narrowest rocking curve FWHM values, indicating substantially improved crystalline quality relative to the on-axis and 2° offcut samples. Previous studies have shown that heteroepitaxial $(\bar{2}01)$ -oriented β-$Ga_2O_3$ films grown on *c*-plane sapphire can exhibit domain tilt, domain-boundary disorder, and rotational domains arising from the symmetry mismatch between monoclinic β-$Ga_2O_3$ and the hexagonal surface symmetry of sapphire. Cui et al. demonstrated that improved epitaxial growth conditions can reduce domain tilt and domain-boundary disorder,[64] while Rafique et al. showed that increasing the sapphire offcut angle favors a single in-plane orientation and promotes step-flow growth.[27] The improved terrace alignment and reduced rocking-curve FWHM observed in the present Ge-doped β-$Ga_2O_3$ films are consistent with reduced rotational-domain and twin-related structural disorder at higher offcut angles.[27,64]

Raman spectroscopy was employed to further assess the phase purity, crystallographic orientation, and lattice quality of the LPCVD-grown Ge-doped β-$Ga_2O_3$ film on a 6° offcut *c*-plane sapphire, as shown in Figure 4. The Raman spectrum of the heteroepitaxial film exhibits a series of well-defined vibrational modes characteristic of β-$Ga_2O_3$, with no additional peaks associated with other $Ga_2O_3$ polymorphs. In addition to the sapphire substrate modes, prominent Raman peaks are observed at approximately 114, 144, 169, 200, 319, 347, 416, 474, 630, 652, and 767 $cm^{-1}$. The peaks at ∼114, 144, 474, and 652 $cm^{-1}$ correspond to $B_g$ vibrational modes, while the remaining peaks are assigned to $A_g$ modes.[27,65] The low-frequency modes below 200 $cm^{-1}$ are associated with lattice translations and vibrations of interconnected $GaO_4$ tetrahedra and $GaO_6$ octahedral chains, the mid-frequency modes between ∼300 and 420 $cm^{-1}$ arise from $GaO_6$ octahedral deformations, and the high-frequency modes above 600 $cm^{-1}$ originate from Ga–O stretching vibrations in $GaO_4$ tetrahedra. The sharp and high-intensity $A_g$ and $B_g$ modes indicate good crystalline order and phase purity of the heteroepitaxial β-$Ga_2O_3$ film, consistent with the narrow XRD rocking-curve widths and smooth surface morphology observed for the 6° offcut sample.

XPS was performed on the representative Ge-doped β-$Ga_2O_3$ film grown on a 6° offcut *c*-plane sapphire to examine the surface chemical composition and bonding states, as shown in Figure 5. The wide-scan survey spectrum in Figure 5a confirms the presence of Ga and O as the dominant constituents, with no detectable unintended impurity-related peaks. High-resolution O 1s spectra (Figure 5b) can be deconvoluted into a dominant Ga–O bonding component centered near ∼530 eV and a weaker higher-binding-energy shoulder attributed to surface hydroxyl (O–H) species. Similarly, the Ga 3d spectrum (Figure 5c) is dominated by the Ga–O bonding component, with a contribution associated with Ga–OH or oxygen-related surface states. Quantitative compositional analysis was carried out using the integrated peak areas corrected by the corresponding relative sensitivity factors. The normalized O 1s and Ga 3d areas yield an oxygen atomic concentration of 59.70% and a gallium atomic concentration of 40.30%, corresponding to an O/Ga ratio of approximately 1.48. This value is close to the ideal stoichiometric ratio of 1.5 for $Ga_2O_3$, indicating a near-stoichiometric composition of the heteroepitaxial film.

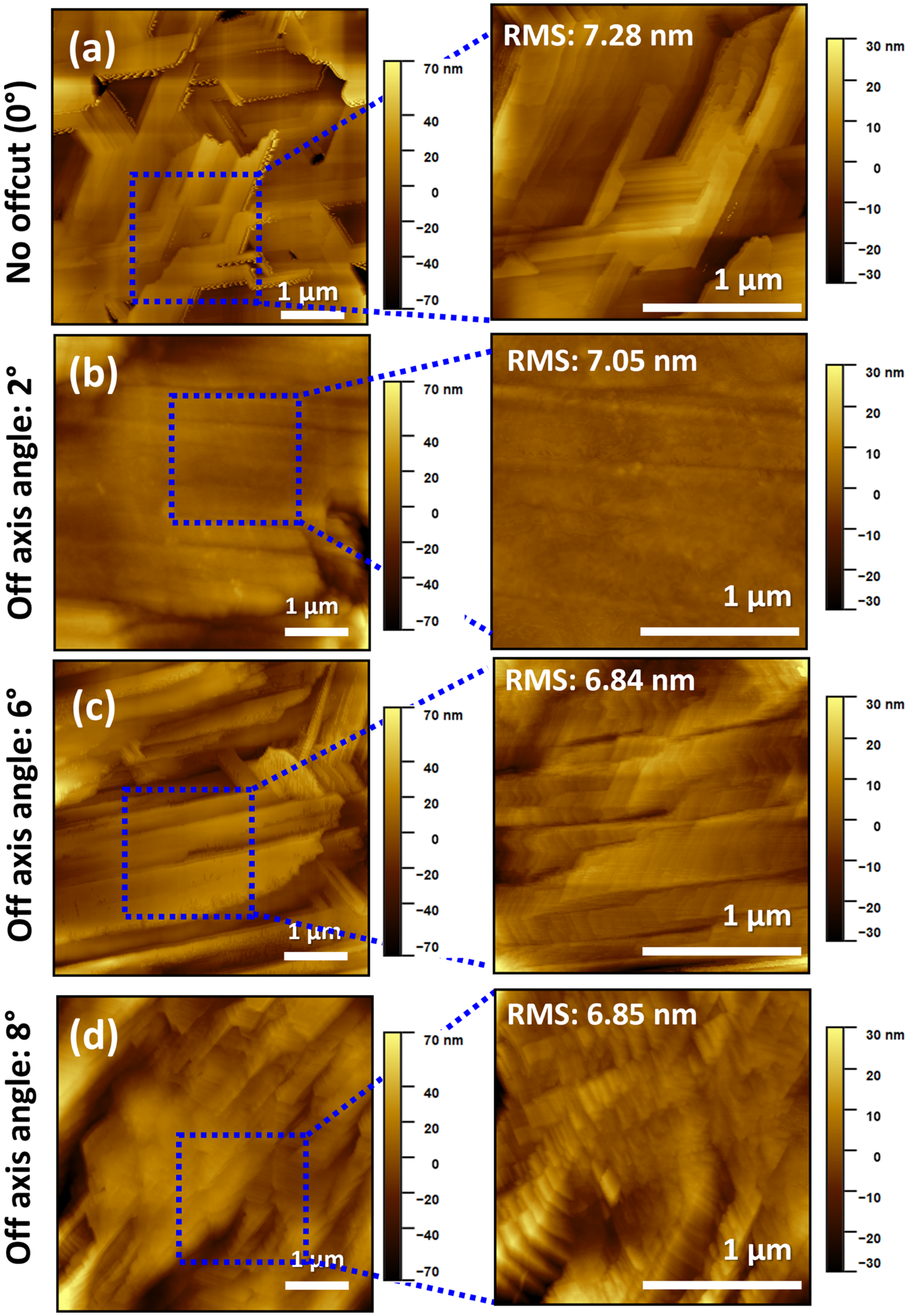

**Figure 2.** Surface AFM images of Ge-doped β-$Ga_2O_3$ thin films heteroepitaxially grown on *c*-plane sapphire substrates with different offcut angles: (a) nominally on-axis (0°), (b) 2°, (c) 6°, and (d) 8° offcut.

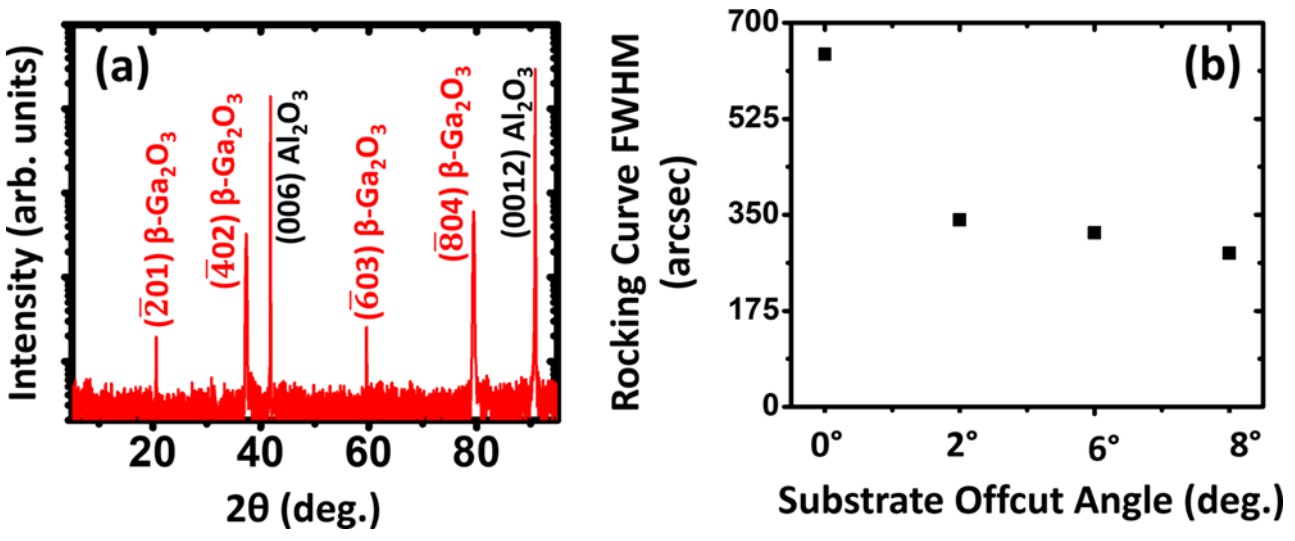

**Figure 3.** (a) High-resolution XRD ω–2θ scan of the Ge-doped β-$Ga_2O_3$ film heteroepitaxially grown on a 6° offcut *c*-plane sapphire substrate, showing strong preferential orientation along the ($\bar{2}$01) family of β-$Ga_2O_3$ planes. (b) Rocking curve FWHM measured from the symmetric ($\bar{4}$02) β-$Ga_2O_3$ reflection as a function of sapphire substrate offcut angle, showing a systematic reduction in rocking curve FWHM with increasing offcut angle.

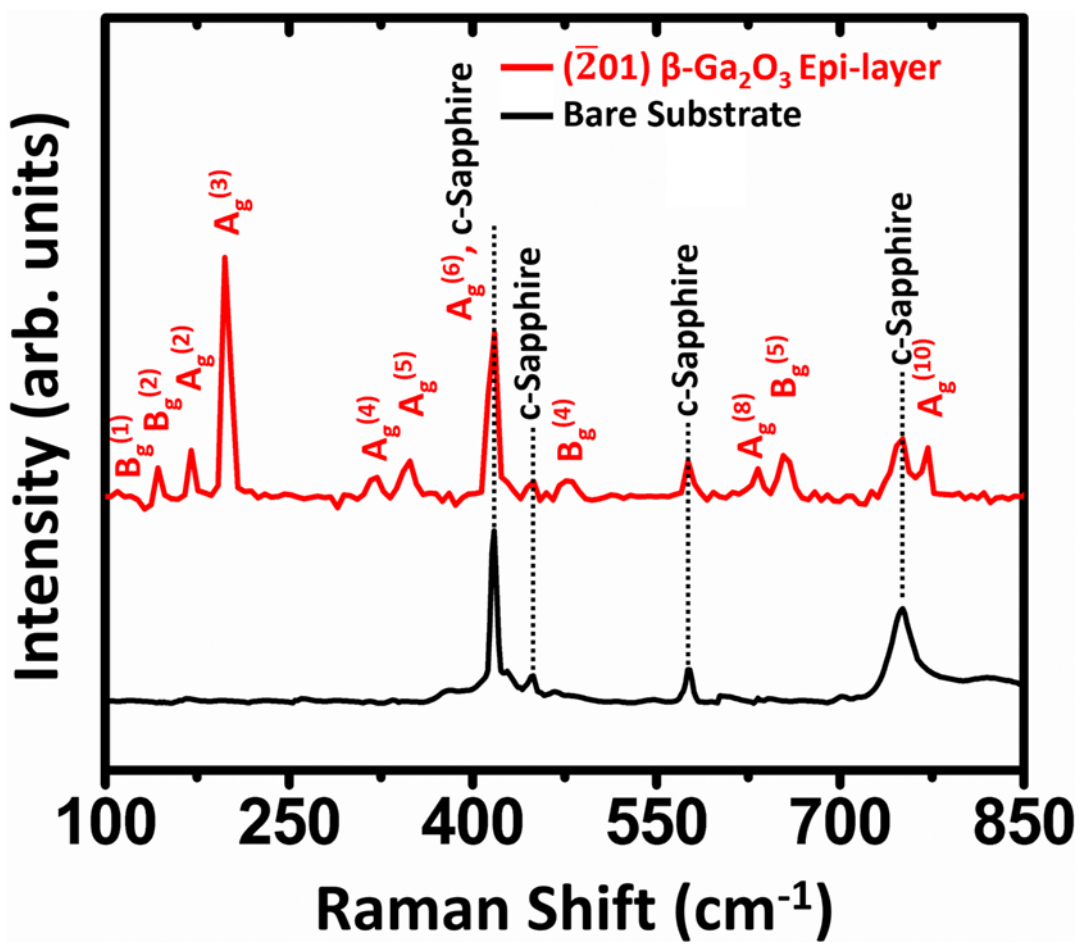

**Figure 4.** Room-temperature Raman spectra of the Ge-doped β-$Ga_2O_3$ heteroepitaxial film on a 6° offcut *c*-plane sapphire substrate, together with the bare sapphire substrate spectrum.

The substrate offcut angle exhibits a pronounced influence on the electron mobility of the Ge-doped β-$Ga_2O_3$ films, as summarized in Table 1. A systematic evolution of the electrical transport properties is observed as the sapphire offcut angle increases from 0° to 8°, indicating that substrate-induced step structure and growth kinetics strongly influence crystalline quality and carrier transport. The measured carrier concentrations span from $1.43 \times 10^{17}$ to $2.75 \times 10^{18}$ $cm^{-3}$. The film grown on nominally on-axis sapphire exhibits a relatively low room-temperature mobility of 15 $cm^2$/V·s at a carrier concentration of $2.45 \times 10^{18}$ $cm^{-3}$, suggesting significant carrier scattering associated with increased structural disorder and less effective step-flow growth. Introducing a 2° offcut results in a substantial mobility enhancement to 47–57 $cm^2$/V·s at carrier concentrations of ~2.5 to $2.8 \times 10^{18}$ $cm^{-3}$, reflecting improved crystalline quality and more efficient adatom incorporation enabled by the presence of atomic step terraces. The most significant mobility enhancement is observed for the 6° offcut samples, which exhibit room-temperature mobilities ranging from 117 to 71 $cm^2$/V·s at corresponding carrier concentrations ranging from $1.43 \times 10^{17}$ to $4.55 \times 10^{17}$ $cm^{-3}$, respectively. This improvement is attributed to enhanced step-flow growth, improved terrace alignment, reduced defect density, and superior crystalline quality, consistent with the SEM, AFM, and XRD results discussed previously. For the 8° offcut samples, the mobility remains relatively high at 64–65 $cm^2$/V·s despite variations in carrier concentration from $3.92 \times 10^{17}$ to $7.42 \times 10^{17}$ $cm^{-3}$. Although the 8° samples maintain highly ordered terrace structures and good crystalline quality, the slightly lower mobility compared to the best-performing 6° sample suggests that the higher step density and localized step bunching associated with larger substrate miscut angles may increase structural disorder and dislocation-related scattering, thereby partially limiting further mobility enhancement.

To place these results in context, Figure 6 compares the room-temperature Hall mobility of the present LPCVD-grown Ge-doped β-$Ga_2O_3$ films with previously reported Ge-doped β-$Ga_2O_3$ grown by MBE,[23] MOCVD,[60] LPCVD,[63] and mist-CVD,[37] as well as Ge-doped α-$Ga_2O_3$ grown by mist-CVD.[66,67] Ge-doped β-$Ga_2O_3$ films grown on native (010) β-$Ga_2O_3$ substrates have previously demonstrated room-temperature mobilities ranging from ~140 to ~38 $cm^2$/V·s over a free carrier concentration range of ~$2 \times 10^{16}$ to ~$3 \times 10^{20}$ $cm^{-3}$, depending on the growth technique and doping level.[23,60] In contrast, previously reported Ge-doped β-$Ga_2O_3$ heteroepitaxial films grown on sapphire generally exhibit substantially lower mobilities,[63] typically not exceeding ∼26 to 28 $cm^2$/V·s. The highest mobility achieved in the present work, 117 $cm^2$/V·s at a carrier concentration of $1.43 \times 10^{17}$ $cm^{-3}$, therefore represents a significant improvement over previously reported Ge-doped β-$Ga_2O_3$ films on sapphire (more than a fourfold higher mobility) and approaches the mobility range commonly associated with homoepitaxial growth on native β-$Ga_2O_3$ substrates. Furthermore, recently reported Ge-doped α-$Ga_2O_3$ films grown by mist-CVD on m-plane sapphire have demonstrated room-temperature mobilities of up to ∼99 $cm^2$/V·s.[66] The mobility achieved in the present work is comparable to the highest reported values for Ge-doped α-$Ga_2O_3$ films, indicating the excellent transport properties obtained through sapphire offcut engineering in LPCVD-grown Ge-doped β-$Ga_2O_3$ heteroepitaxy.

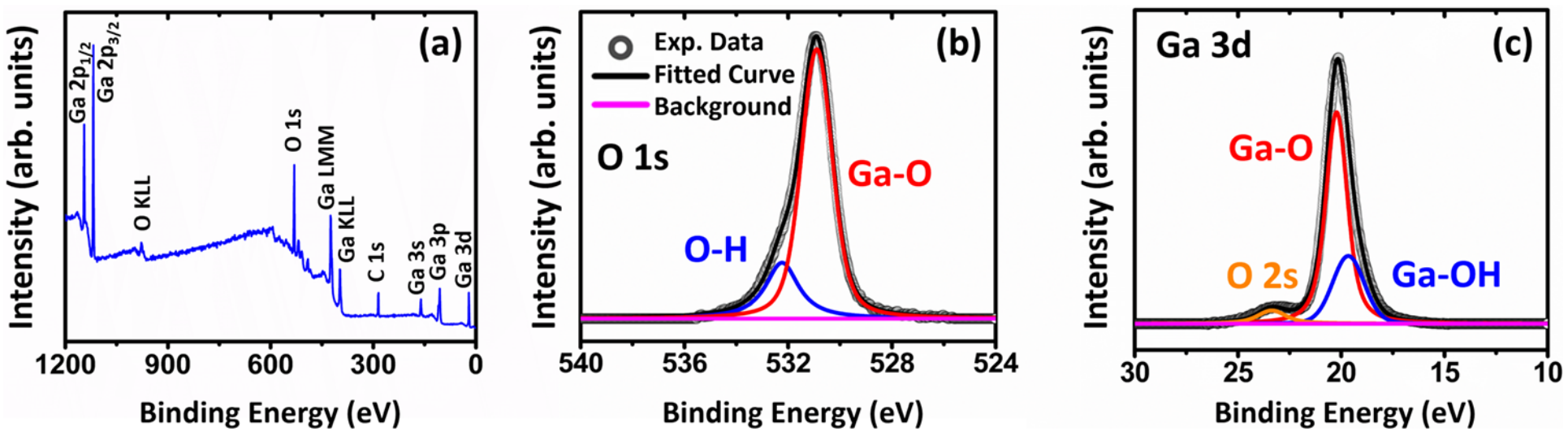

**Figure 5.** XPS analysis of the Ge-doped β-$Ga_2O_3$ heteroepitaxial film on a 6° offcut *c*-plane sapphire substrate. (a) Wide-scan XPS survey spectrum confirming the presence of Ga and O elemental constituents with no detectable impurity-related peaks. (b) High-resolution O 1s spectrum deconvoluted into the dominant Ga–O bonding component and a higher-binding-energy shoulder associated with surface hydroxyl (O–H) species. (c) High-resolution Ga 3d spectrum showing the primary Ga–O bonding contribution together with a Ga–OH component.

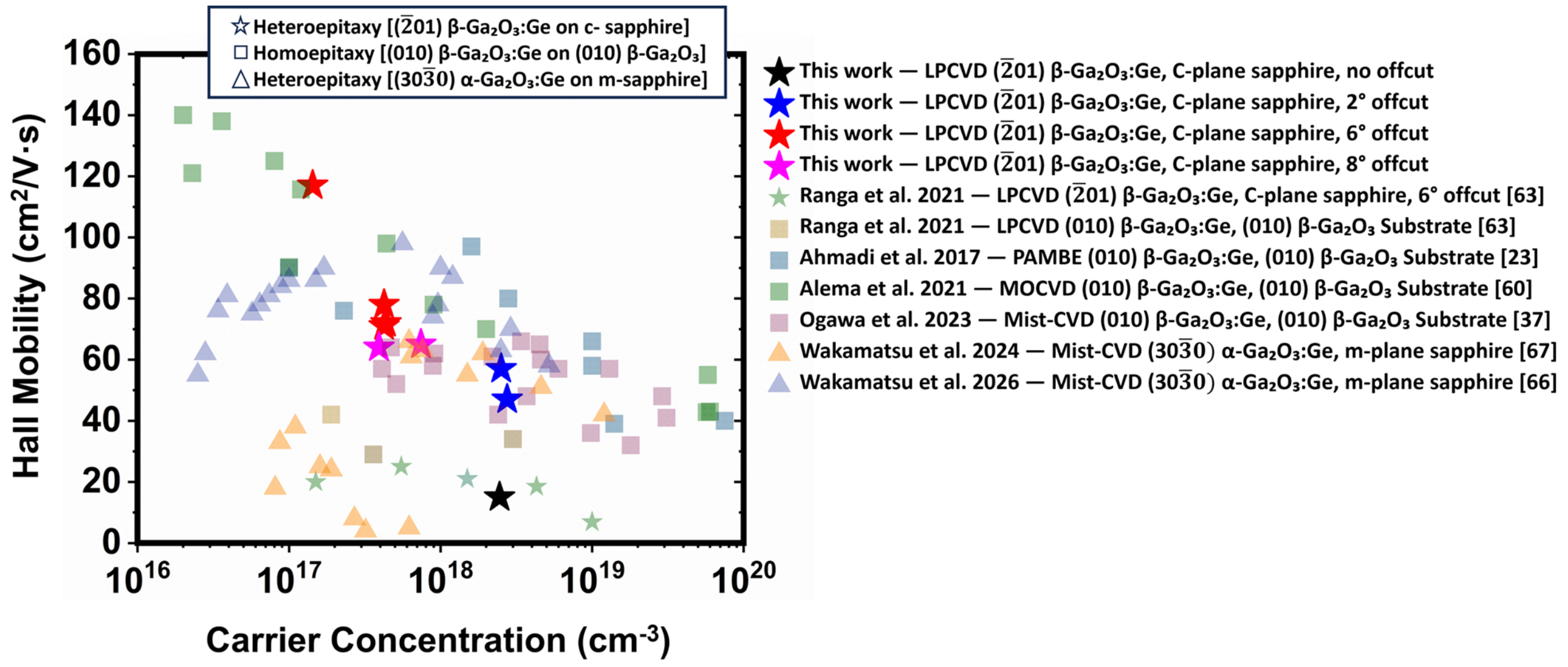

**Figure 6.** Benchmark comparison of room-temperature Hall mobility as a function of carrier concentration for Ge-doped β-$Ga_2O_3$ and α-$Ga_2O_3$ films grown by different epitaxial techniques and on different substrates.

To further probe the dominant scattering mechanisms and assess the transport behavior beyond room temperature, temperature-dependent Hall measurements were carried out on Ge-doped β-$Ga_2O_3$ films grown on sapphire substrates with different offcut angles. Figure 7 shows the temperature dependence of the electron concentration and Hall mobility for Ge-doped β-$Ga_2O_3$ films grown on sapphire substrates with different offcut angles. For the film grown on on-axis sapphire (Sample No. 1), the room-temperature carrier concentration is $2.45 \times 10^{18}$ $cm^{-3}$ with a low mobility of 15 $cm^2$/V·s, which increases only modestly to a peak mobility of 17 $cm^2$/V·s at 192 K. This weak temperature dependence indicates strong defect-related scattering that dominates transport across the measured temperature range. Introducing a 2° offcut leads to an improvement in transport. The room-temperature mobility increases to 47 $cm^2$/V·s at a carrier concentration of $2.75 \times 10^{18}$ $cm^{-3}$ (Sample No. 3), while the peak mobility reaches 68 $cm^2$/V·s at 148 K with a corresponding carrier concentration of $2.31 \times 10^{18}$ $cm^{-3}$. The moderate increase in mobility upon cooling suggests partial suppression of phonon scattering, although impurity and defect scattering remain influential at these high doping levels. A pronounced improvement is observed for films grown on a 6° offcut sapphire (Sample No. 4). At room temperature, a mobility of 117 $cm^2$/V·s is achieved at a carrier concentration of $1.43 \times 10^{17}$ $cm^{-3}$. Upon cooling, the mobility increases sharply to a peak value of 337 $cm^2$/V·s at 128 K, while the carrier concentration decreases to $8.96 \times 10^{16}$ $cm^{-3}$. The strong temperature dependence and high peak mobility indicate a reduction in defect-related scattering. Notably, the room-temperature mobility of 117 $cm^2$/V·s and the peak low-temperature mobility of 337 $cm^2$/V·s represent the highest reported values for Ge-doped β-$Ga_2O_3$ films grown on sapphire substrates. These values substantially exceed those previously reported for Ge-doped β-$Ga_2O_3$ heteroepitaxial films on sapphire and also surpass the mobility reported for Ge-doped β-$Ga_2O_3$ homoepitaxial films grown on native (010) β-$Ga_2O_3$ substrates by LPCVD.[63] For the 8° offcut sample (Sample No. 8), the room-temperature mobility is 64 $cm^2$/V·s at a comparatively low carrier concentration of $3.92 \times 10^{17}$ $cm^{-3}$. The mobility increases further to a peak value of 212 $cm^2$/V·s at 128 K, accompanied by a reduction in carrier concentration to $1.64 \times 10^{17}$ $cm^{-3}$.

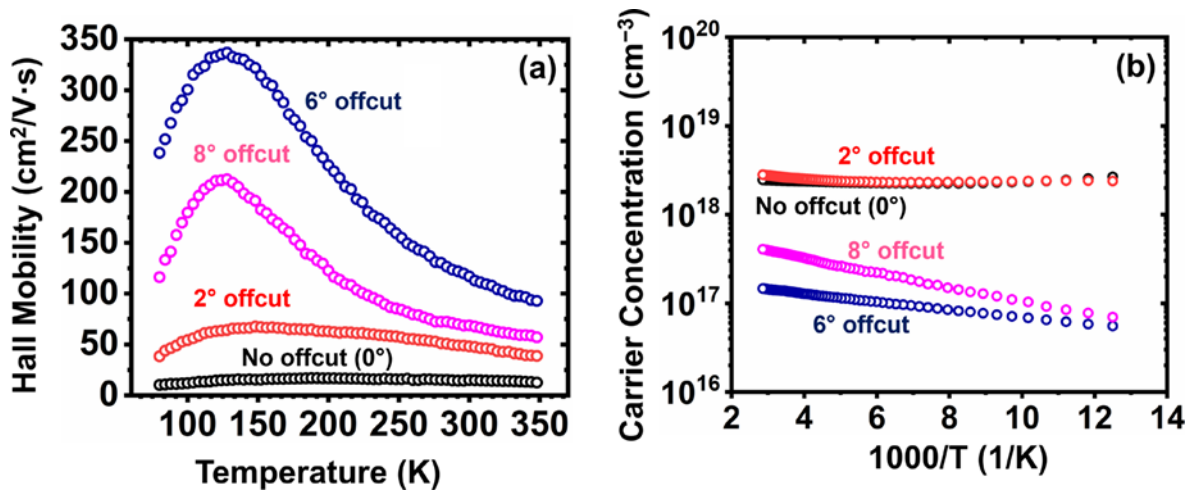

**Figure 7.** Temperature-dependent Hall transport characteristics of Ge-doped β-$Ga_2O_3$ heteroepitaxial films on *c*-plane sapphire substrates with different offcut angles. (a) Hall mobility as a function of temperature for films grown on nominally on-axis (0°), 2°, 6°, and 8° offcut sapphire substrates. (b) Temperature-dependent carrier concentration plotted as a function of 1000/*T*.

The influence of carrier concentration on electron transport was further examined using temperature-dependent Hall measurements on two representative 6° offcut samples with different doping levels, as shown in Figure 8. Sample 4, with a room-temperature carrier concentration of $1.43 \times 10^{17}$ cm$^{-3}$, exhibits a room-temperature mobility of 117 cm$^2$/V·s and a peak mobility of 337 cm$^2$/V·s at 128 K. Sample 5, with a relatively higher carrier concentration of $4.24 \times 10^{17}$ cm$^{-3}$, exhibits a room-temperature mobility of 78 cm$^2$/V·s and a peak mobility of 149 cm$^2$/V·s at 124 K. The higher mobility observed for the lower-doped sample, while maintaining the same 6° sapphire offcut condition, indicates that carrier concentration plays a significant role in determining electron transport properties. The lower-doped sample maintains a lower electron concentration throughout the measured temperature range, which is consistent with reduced ionized impurity scattering and consequently enhanced electron mobility.

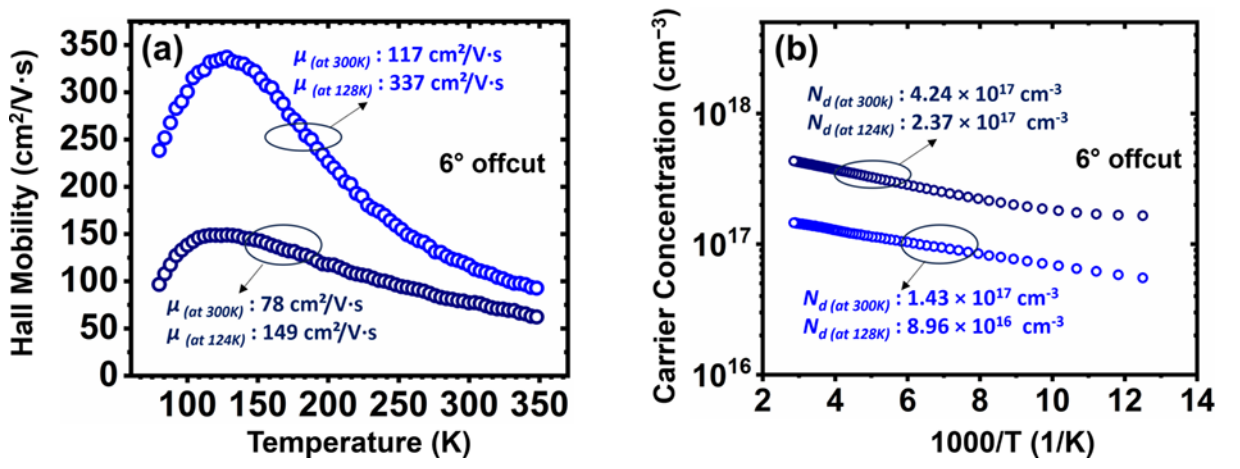

**Figure 8.** Temperature-dependent Hall transport characteristics of two representative Ge-doped β-$Ga_2O_3$ films grown on 6° offcut *c*-plane sapphire. (a) Hall mobility as a function of temperature for Sample No. 4 and Sample No. 5. (b) Corresponding carrier concentration as a function of 1000/*T*.

To further probe the carrier transport behavior and identify the dominant scattering mechanisms influencing electron mobility, temperature-dependent Hall-effect data for the 6° and 8° offcut sapphire samples were analyzed using a charge-neutrality model combined with Boltzmann transport calculations.[66,68,69] The experimentally measured temperature dependences of carrier concentration and Hall mobility, together with the corresponding fitted curves, are shown in Figure 9. The fitting incorporates multiple carrier scattering contributions to evaluate their relative influence on the temperature-dependent transport characteristics. The temperature dependence of the electron concentration $n(T)$ was analyzed using a charge-neutrality condition that accounts for ionized donors and compensating acceptor-like centers. In heteroepitaxial β-$Ga_2O_3$ films, threading dislocations are known to introduce acceptor-like traps along the dislocation core, which capture free electrons and contribute to compensation. Accordingly, the charge-neutrality condition can be written as

$$n(T) + N_A + \left(\frac{N_{dis}}{d}\right) = N_{D1}^+(T) + N_{D2}^+(T) \quad (1)$$

where $N_A$ is the bulk acceptor concentration, $N_{dis}$ is the threading dislocation density, and $d$ is the spacing between acceptor centers along the dislocation line, approximated using the crystallographic periodicity along the $(\bar{2}01)$ growth direction,[70,71] yielding $d \approx 0.47$ nm. $N_{D1}^+$ and $N_{D2}^+$ represent two distinct donor populations with activation energies $E_{D1}$ and $E_{D2}$, respectively. The ionized fraction of each donor is described by Fermi–Dirac statistics as

$$N_{Di}^+(T) = \frac{N_{Di}}{1 + 2\exp\left(\frac{-(E_{Di} - E_F)}{k_B T}\right)} \quad (2)$$

where $E_F$ is the Fermi level and $k_B$ is Boltzmann's constant. Equations 1 and 2 were solved self-consistently to fit the experimentally measured $n(T)$.

The extracted fitting parameters are summarized in Table 2. For the 6° offcut sample (Sample No. 4), the fitting yields donor activation energies of 12.5 and 80 meV, together with an acceptor concentration of $2 \times 10^{15}$ cm$^{-3}$ and a threading dislocation density of $1.0 \times 10^9$ cm$^{-2}$. For the 8° offcut sample (Sample No. 8), the extracted donor activation energies are 19 and 80 meV, with an acceptor concentration of $5 \times 10^{15}$ cm$^{-3}$ and a threading dislocation density of $1.7 \times 10^9$ cm$^{-2}$. In both cases, the shallow donor population primarily contributes to room-temperature transport, while the deeper donor state becomes increasingly activated at elevated temperatures. Similar multi-donor behavior has been widely reported in β-$Ga_2O_3$ and is commonly associated with native defects and unintentional impurities.[2,4,72] The extracted acceptor concentration remains low, indicating weak compensation in both films. The temperature-dependent Hall mobility, $\mu(T)$, was analyzed by considering multiple scattering mechanisms. The total mobility is calculated using Matthiessen's rule[68,73] as

$$\frac{1}{\mu_{total}(T)} = \frac{1}{\mu_{POP}(T)} + \frac{1}{\mu_{II}(T)} + \frac{1}{\mu_{NI}(T)} + \frac{1}{\mu_{ADP}(T)} + \frac{1}{\mu_{DIS}(T)} \quad (3)$$

where $\mu_{POP}$, $\mu_{II}$, $\mu_{NI}$, $\mu_{ADP}$, and $\mu_{DIS}$ represent the mobility contributions limited by polar optical phonon scattering, ionized impurity scattering, neutral impurity scattering, acoustic deformation potential scattering, and dislocation scattering, respectively. The material parameters used in the calculations are summarized in Table 3 and were selected based on previously reported theoretical and experimental studies on β-$Ga_2O_3$ transport properties.[2,4,72] The calculated mobility behavior indicates that polar optical phonon scattering dominates at elevated temperatures, whereas ionized impurity, neutral impurity, and dislocation scattering become increasingly important at lower temperatures. The 6° offcut sample exhibits the highest measured mobility and the lowest extracted threading dislocation density ($1.0 \times 10^9$ cm$^{-2}$), while the 8° offcut sample exhibits a higher dislocation density of $1.7 \times 10^9$ cm$^{-2}$. These values are comparable to previously reported threading dislocation densities for heteroepitaxial $Ga_2O_3$ films grown on lattice-mismatched substrates.[66,74] Overall, the agreement between the modeled and experimentally measured carrier concentration and mobility indicates that the adopted charge-neutrality and mobility-scattering model captures the dominant transport processes in the present Ge-doped β-$Ga_2O_3$ heteroepitaxial films.

**Table 2. Extracted Donor Concentrations, Donor Activation Energies, Acceptor Concentrations, and Threading Dislocation Densities Obtained from Fitting the Temperature-Dependent Hall-Effect Data for the 6° and 8° Offcut Ge-Doped β-$Ga_2O_3$ Films**

| sample no. | substrate miscut (deg) | $N_{D1}$ (cm$^{-3}$) | $E_{D1}$ (meV) | $N_{D2}$ (cm$^{-3}$) | $E_{D2}$ (meV) | $N_A$ (cm$^{-3}$) | $N_{dis}$ (cm$^{-2}$) |
|---|---|---|---|---|---|---|---|
| 4 | 6 | $1.59 \times 10^{17}$ | 12.5 | $5.0 \times 10^{16}$ | 80 | $2 \times 10^{15}$ | $1.0 \times 10^9$ |
| 8 | 8 | $4.98 \times 10^{17}$ | 19 | $1.7 \times 10^{17}$ | 80 | $5 \times 10^{15}$ | $1.7 \times 10^9$ |

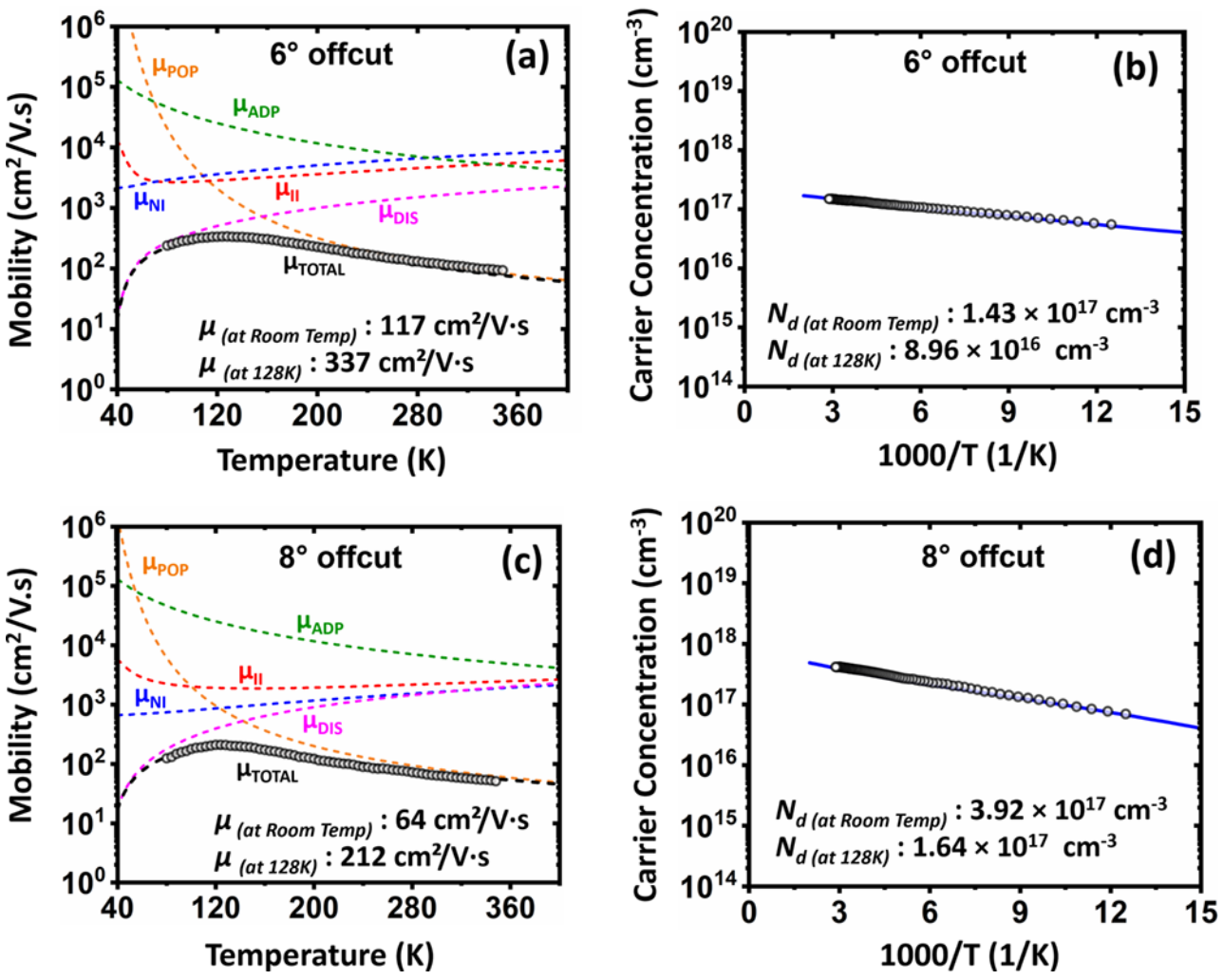

**Figure 9.** Temperature-dependent transport modeling of Ge-doped β-$Ga_2O_3$ heteroepitaxial films grown on (a, b) 6° and (c, d) 8° offcut *c*-plane sapphire substrates. (a, c) Calculated mobility contributions from polar optical phonon scattering ($\mu_{POP}$), acoustic deformation potential scattering ($\mu_{ADP}$), ionized impurity scattering ($\mu_{II}$), neutral impurity scattering ($\mu_{NI}$), and dislocation scattering ($\mu_{DIS}$), together with the total modeled mobility ($\mu_{TOTAL}$). The measured Hall mobility data are shown by the open circles. (b, d) Temperature-dependent carrier concentration as a function of 1000/*T* together with the fitted charge-neutrality model. The modeled results show good agreement with the experimental data for both offcut conditions.

## 4. CONCLUSIONS

In summary, high-quality Ge-doped ($\bar{2}01$) β-$Ga_2O_3$ thin films were successfully grown on *c*-plane sapphire substrates with different offcut angles using LPCVD. Systematic structural, morphological, chemical, and electrical characterization revealed that the sapphire substrate offcut plays a critical role in governing the growth mode, crystalline quality, and carrier transport behavior of the heteroepitaxial β-$Ga_2O_3$ films. SEM and AFM analyses demonstrated a clear transition from irregular multidirectional growth on nominally on-axis substrates to highly ordered and continuous terrace structures at higher offcut angles, indicating the promotion of step-flow growth with increasing substrate miscut. The films grown on 6° and 8° offcut sapphire substrates exhibited significantly improved terrace alignment, reduced surface roughness, and enhanced lateral surface continuity. High-resolution XRD measurements confirmed phase-pure monoclinic β-$Ga_2O_3$ growth with strong preferential orientation along the ($\bar{2}01$) family of planes. The reduction in rocking-curve FWHM with increasing offcut angle further demonstrated a substantial improvement in crystalline coherence and structural quality. Raman spectroscopy verified the phase purity and high crystalline order of the heteroepitaxial films. XPS analysis further confirmed a near-stoichiometric film composition with an O/Ga ratio close to the ideal value for $Ga_2O_3$ and no detectable impurity-related phases. Electrical transport measurements revealed a strong dependence of carrier mobility on the substrate offcut angle. The 6° offcut sample exhibited a room-temperature mobility of 117 cm$^2$/V·s at a carrier concentration of $1.43 \times 10^{17}$ cm$^{-3}$ and a peak low-temperature mobility of 337 cm$^2$/V·s at 128 K, representing the highest reported room-temperature and low-temperature mobilities for Ge-doped β-$Ga_2O_3$ films grown on sapphire substrates. The improved transport behavior observed with increasing sapphire offcut angle is attributed to the combined effects of enhanced step-flow growth, improved crystalline quality, reduced structural disorder, and reduced dislocation-related carrier scattering. To further understand the transport behavior, temperature-dependent carrier concentration and mobility were analyzed using charge-neutrality and mobility-scattering models. The fitting results revealed two donor populations with activation energies of 12.5–19 and 80 meV, respectively, together with relatively low acceptor concentrations. The extracted threading dislocation densities for the 6° and 8° offcut samples were $1.0 \times 10^9$ and $1.7 \times 10^9$ cm$^{-2}$, respectively. Overall, this work demonstrates that substrate offcut engineering is an effective approach for improving the structural and electrical properties of LPCVD-grown heteroepitaxial Ge-doped β-$Ga_2O_3$ films on sapphire. The observed enhancement in crystalline quality and carrier mobility with optimized sapphire offcut angles provides important insights into heteroepitaxial growth kinetics and carrier transport mechanisms, contributing to the continued development of high-performance β-$Ga_2O_3$ power electronic devices on scalable foreign substrates.

**Table 3. Material Parameters Used for Charge-Neutrality and Mobility Calculations in Ge-Doped β-$Ga_2O_3$ Films**

| calculation parameter | symbol | value |
|---|---|---|
| phonon energy (meV) | $\hbar\omega_0$ | 47 (6° offcut) and 39 (8° offcut) (fitted) |
| dielectric constant | $\varepsilon_s$ | 10.2 |
| high-frequency dielectric constant | $\varepsilon_{s,\infty}$ | 3.6 |
| electron effective mass ($m_0$) | $m^*$ | 0.313 |
| acoustic deformation potential (eV) | $E_{ADP}$ | 6.9 |
| mass density (kg·m$^{-3}$) | $\rho$ | $5.88 \times 10^3$ |
| sound velocity (m·s$^{-1}$) | $v_s$ | $6.8 \times 10^3$ |

## ASSOCIATED CONTENT

### Data Availability Statement

The data that support the findings of this study are available from the corresponding author upon reasonable request.

## AUTHOR INFORMATION

### Corresponding Author

**Anhar Bhuiyan** – *Department of Electrical and Computer Engineering, University of Massachusetts Lowell, Lowell, Massachusetts 01854, United States;* orcid.org/0000-0003-3421-2813; Email: anhar_bhuiyan@uml.edu

### Authors

**Ahmed Ibreljic** – *Department of Electrical and Computer Engineering, University of Massachusetts Lowell, Lowell, Massachusetts 01854, United States*

**Saleh Ahmed Khan** – *Department of Electrical and Computer Engineering, University of Massachusetts Lowell, Lowell, Massachusetts 01854, United States*

**Sourav Sarker** – *Department of Electrical and Computer Engineering, University of Massachusetts Lowell, Lowell, Massachusetts 01854, United States*

**Ibrahim Isah** – *Department of Chemical Engineering, University of Massachusetts Lowell, Lowell, Massachusetts 01854, United States*

**Stephen Margiotta** – *Department of Electrical and Computer Engineering, University of Massachusetts Lowell, Lowell, Massachusetts 01854, United States*

**Michael T. Davenport** – *Department of Chemical Engineering, University of Massachusetts Lowell, Lowell, Massachusetts 01854, United States*

**Stephen Lam** – *Department of Chemical Engineering, University of Massachusetts Lowell, Lowell, Massachusetts 01854, United States;* orcid.org/0000-0002-7683-1201

Complete contact information is available at: https://pubs.acs.org/doi/10.1021/acsaelm.6c01043

### Notes

The authors declare no competing financial interest.

## ACKNOWLEDGEMENTS

The authors acknowledge the funding support from the National Science Foundation (NSF) under award numbers ECCS-2532898 and ECCS-2501623. M. Davenport and S. Lam acknowledge support from the U.S. Nuclear Regulatory Commission (NRC) under Grant No. 31310024M0046.